\documentclass{article}
\usepackage{ijcai20}
\usepackage[backend=biber]{biblatex}
\bibliography{ijcai20}

\usepackage{times}

\usepackage{soul}
\usepackage{url}
\usepackage[utf8]{inputenc}
\usepackage[small]{caption}
\usepackage{graphicx}
\usepackage{amsmath}
\usepackage{booktabs}
\usepackage{subfigure}
\title{Missed calls, Automated Calls and Health Support: Using AI to improve maternal health outcomes by increasing program engagement}

\author{
Siddharth Nishtala$^{1}$\footnote{Contact Author}\and
Harshavardhan Kamarthi$^{1}$\and
Divy Thakkar$^2$\and\\
Dhyanesh Narayanan$^2$\and
Anirudh Grama$^2$\and
Aparna Hegde$^3$\and
Ramesh Padmanabhan$^3$\and\\
Neha Madhiwalla$^3$\and
Suresh Chaudhary$^3$\and
Balaraman Ravindran$^{1}$\and
Milind Tambe$^2$
\affiliations
$^1$Robert Bosch Centre for Data Science and Artificial Intelligence\\
Indian Institute of Technology, Madras\\
$^2$Google Research, India\\
$^3$ARMMAN\\
}

\begin{document}

\maketitle

\begin{abstract}

India accounts for 11\% of maternal deaths globally where a woman dies in childbirth every fifteen minutes. Lack of access to preventive care information is a significant problem contributing to high maternal morbidity and mortality numbers, especially in low-income households. We work with ARMMAN, a non-profit based in India, to further the use of call-based information programs by early-on identifying women who might not engage on these programs that are proven to affect health parameters positively. 
We analyzed anonymized call-records of over 300,000 women registered in an awareness program created by ARMMAN that uses cellphone calls to regularly disseminate health related information. We built robust deep learning based models to predict short term and long term dropout risk from call logs and beneficiaries' demographic information. Our model performs 13\% better than competitive baselines for short-term forecasting and 7\% better for long term forecasting. We also discuss the applicability of this method in the real world through a pilot validation that uses our method to perform targeted interventions.

\end{abstract}

\section{Introduction}

In a 2017 report on maternal mortality \cite{10665-327596}, the World Health Organization (WHO) estimated a total of 295,000 deaths worldwide and India accounts for nearly 12\% of these deaths. India, with a population of over a billion, has an estimated maternal mortality ratio of 145 (maternal deaths per 100,000 livebirths). Majority of these deaths are preventable and are accounted to information asymmetry. There is immense potential to leverage ICTs to mitigate pregnancy related risks and improve maternal health by closing information gaps. 

One of the major contributory factors for preventable maternal deaths is delay in getting adequate care. The three delay model \cite{thaddeus1994too} proposes that maternal mortality is associated with three phases of delay. The first is in deciding to seek care which is influenced by various factors involved in decision-making such as socio-cultural factors, individual or family's understanding of complications and risk factors in pregnancy, and financial implications. The second delay is in physically reaching an adequate healthcare facility due to availability and accessibility constraints.  The third delay is associated with the availability of trained healthcare professionals, supplies and equipment to provide adequate care. 

Socio-economic context of a person plays a critical role in determining maternal and child mortality parameters. The decision to seek care could be significantly impacted with a broadened access to information on reproductive health. Many organisations have begun using digital tools for improving health related awareness and care to pregnant women. \textit{ARMMAN}, a non-profit organization based in Mumbai works on bridging this gap through their program \textit{mMitra}, which has catered to over 2 million women. \textit{mMitra} is a free mobile voice call service that sends timed and targeted preventive care information directly to the phones of the enrolled women in their chosen language and time slot. A series of 141 individualised voice messages of 60-120 seconds are sent with different frequencies at different stages throughout the course of pregnancy and up to 12 months after child birth. 

The program witnesses a significant dropout rate  owing to a number of reasons grounded in their socio-cultural context through community health practices, lack of information, misinformation, etc.
Programs such as mMitra strive to increase maternal health parameters and we recognize the value of identifying and providing timely and relevant interventions to assure program retention.
Interventions include providing reminders through missed calls, personalized health consultation based on the projected risk of dropping out. Given the limited resources with non-profit organizations such as \textit{ARMMAN}, there is value in providing predictive modeling services to identify women who have a higher chance of dropping off from these health programs. 

Our contributions are two fold. First, we consider the problem of forecasting women dropping out based on past engagement with the voice call program. We formulate this task as a supervised learning problem and build deep learning based models that are trained on previous call-related data of over 300,000 beneficiaries to accurately predict the risk of disengagement. 
Second, we discuss the deployment of our models as a pilot study to show the effectiveness of our methods in the field.

\section{Related Work}

Predicting customer engagement is beneficial to organizations across a wide range of industries. More commonly known as churn prediction, the problem has been well researched in the fields of telecommunications \cite{HUANG20121414, HUNG2006515}, banking \cite{XIE20095445} and gaming \cite{5284154, runge2014churn}. Most methods typically use customers' demographic or behavioural information to estimate the likelihood of disengagement. Identifying such customers and conducting relevant interventions increases customer retention.

In the context of healthcare, the churn prediction problem, which takes the form of adherence, is well studied for a variety of diseases \cite{10.3389/fphar.2013.00091}. Medication and treatment adherence has been studied for conditions such as heart failure \cite{son2010application}, schizophrenia \cite{howes2012predicting}, HIV \cite{HIV}, Tuberculosis \cite{10.1001/archinte.1996.00440020063008, Killian_2019} etc.. However, most studies first collect data on adherence and the outcome of treatments and then use machine learning techniques to predict outcomes using the collected data. While such studies have helped identify important predictors that influence outcomes, they do not assess the risk of patients during the treatment. To prospectively identify patients that are likely to miss doses, one study trains deep learning models on demographic and past dose information of patients to accurately predict missed doses \cite{Killian_2019} and treatment outcomes. While we use similar methods to predict beneficiaries that are the risk of dropping out, there are a few notable differences in the tasks: (1) outcome prediction in adherence research uses well-defined treatment outcomes, which are not be available in call-based health programs; (2) an additional layer of complexity is introduced as calls sent to a beneficiary might fail for multiple reasons.

\section{Data Description}
During registration of women to the \textit{mMitra} program, demographic data such as age, education level and income group is collected. Further, during the course of the program, every call to a beneficiary is logged.
The dataset collected by \textit{ARMMAN} contained data for all the beneficiaries registered in 2018 and their call records until October 2019. This includes 70,272,621 call records for 329,489 women. Given the sensitivity of the healthcare domain, all the data analyzed was fully anonymized before we further processed it. The important features of the data considered for our work are discussed below.

\textbf{Beneficiary details.} This is collected at the time of registration from all beneficiaries. The important demographic information we considered include age, education level, and income group. We also used other information such as registration date, gestation age at the time of registration, preferred call slot of the day, language of communication and the owner of the phone.

\textbf{Call logs.} The call logs contain the details for all the calls made to every registered beneficiary. Each entry has details of an individual call: message ID, date of call, call duration and information regarding whether the call was successfully connected. As the message content of each call depends on the beneficiary's gestation age at the time, we use the message ID as a proxy for the beneficiary's gestation age.

It is important to note that all the calls attempted to a beneficiary do not necessarily go through. Connection failure can be attributed to two causes: (1) network failure; and (2) if a beneficiary fails to answer the call. While this information on the cause of connection failure could be beneficial for building predictive models, it is not available in the call records. Another feature of the dataset that contributes to its volume is that there are up to two retries for every call that results in a connection failure. We consider the best outcome (longest call duration) for each message. This results in a reduction in size to 27,488,890 call records.

For all further discussion, we will use the following terminology. \texttt{Attempt} is when a call is attempted to a beneficiary. When an attempted call is answered by the beneficiary, it is called a \texttt{connection}. A connection with call duration of at least 30 seconds is called an \texttt{engagement}.

\section{Short term drop-off prediction}

Tracking women who are likely to disengage for an extended period of time based on their past engagement behaviour is very important in mitigating the risk of dropping off. The health care worker can intervene with women who are at risk of dropping off the program by providing reminders in the form of phone calls or by even reaching out to them individually. We, therefore, build models that predict if a woman would not answer any calls for the next two weeks given the call history of past 4 weeks and demographic information.

\paragraph{Problem Formulation}

For each beneficiary registered in the program, we randomly sample call sequences spanning six weeks to build our dataset. The data from the first four weeks are used as input features and data from the final two weeks are used to generate the label. If there are no engagements in the two week prediction period, we label the data as \texttt{high risk}, else it is labelled as \texttt{low risk}. To generate a batch for one training iteration, we sample these sequences from the call logs dataset. We removed the beneficiaries whose demographic features were not complete or were entered in an incorrect format. Additionally, we computed six features from the call logs: number of past attempts, number of past connections, number of past engagements, days since last attempt, days since last connection and days since last engagement. 

\begin{figure}[h]
    \centering
    \includegraphics[width=.95\linewidth]{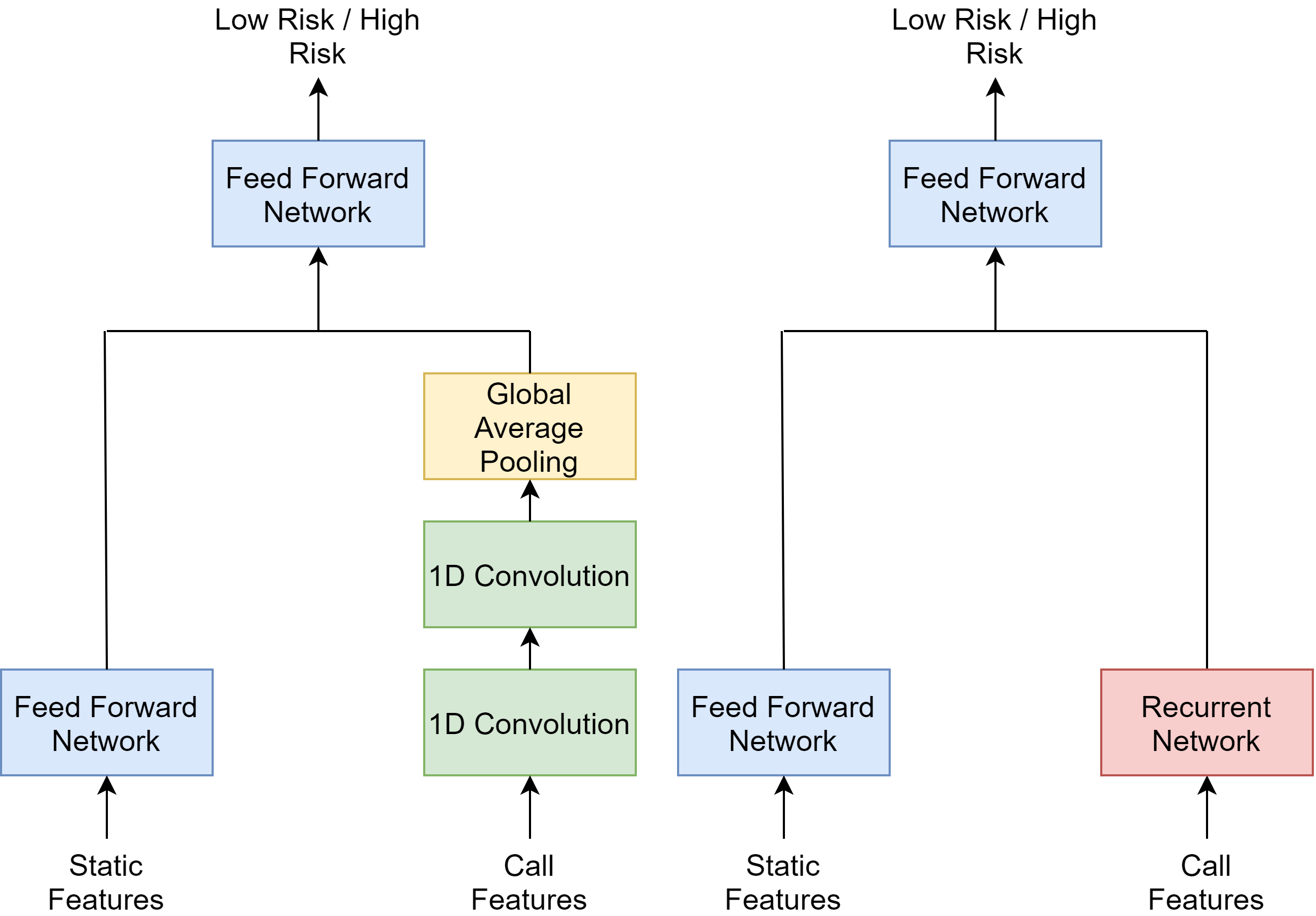}
    \caption{CoNDiP (left) and ReNDiP (right) Architectures}
    \label{fig:ConDiP1}
\end{figure}

\paragraph{Models}

We used random forest model over static demographic features as the baseline model. The random forest model consisted of 200 trees with maximum depth 10. 
In order to fully utilize the static demographic features and the available time-series data of the call histories, we build two neural networks called \textbf{CoNDiP} (Convolutional Neural Disengagement Predictor) and \textbf{ReNDiP} (Recurrent Neural Disengagement Predictor). 

CoNDiP uses 1D Convolutional \cite{kiranyaz20191d} layers to encode the dynamically sized call history features. Since, over the span of 4 weeks at most 8 calls are made to a beneficiary (atmost 2 calls per week), we append call sequences with fewer calls with zeroes and feed it to the 1D convolutional layers. 

In CoNDiP, each convolutional layer has 20 kernels of size 3 and there are two convolutional layer after which we do average pooling as follows. Let $f_{i,j}$ be the value of $i$th feature of $j$th call in the output from the 1D convolutional layers. We do average pooling of these features across time to get features $\hat{f}_{i}=\frac{\sum_{j\in [T] }f_{i,j}}{T}$.

We also encode the static features via two feed forward layers with 50 and 100 neurons. Let $g$ be vector of encoded static features output by the feed forward layer. Then, we concatenate $g$ and $\hat{f}$ and feed it through two feed forward layers each with 100 neurons followed by a final layer to compute the probability of being at risk. After each of the feed-forward layers (except last one), we applied batch normalization. 

We also experiment with encoding time-series data with recurrent layers instead of convolutional layers. By replacing the 1D convolutional layers and average pooling with an LSTM \cite{hochreiter1997long} whose output from the last time-step is taken as input to final feed forward layer, we get \textbf{ReNDiP} (Recurrent Neural Disengagement Predictor). We used a bi-directional LSTM with 100 hidden units. The architectures of CoNDiP and ReNDiP are shown in Figure \ref{fig:ConDiP1}.

\begin{table}[h]
\centering
\begin{tabular}{|l|l|l|l|l|}
\hline
Model        & Accuracy & Precision & Recall & F1 score \\ \hline
RF & 0.70     & 0.71       & 0.72   & 0.71     \\ \hline
CoNDiP       & 0.83     & 0.84      & 0.85   & 0.83     \\ \hline
ReNDiP       & 0.81     & 0.83      & 0.83   & 0.81     \\ \hline
\end{tabular}
\caption{Results for Short-term engagement task. RF: Random Forest}
\label{tab:short-term}
\end{table}

\begin{figure}[h]
    \centering
    \includegraphics[width=.8\linewidth]{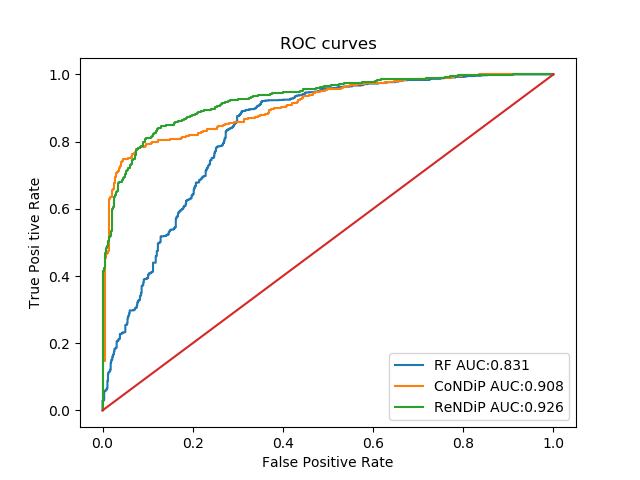}
    \caption{ROC curve and area under the curve (in legend) for the models for short-term engagement task}
    \label{fig:short-term}
\end{figure}
\paragraph{Results}
The results of our models and baseline are summarized in Table \ref{tab:short-term}. CoNDiP and ReNDiP show similar performance scores and clearly outperform the random forest baseline. Due to slightly higher scores, CoNDiP is the model best suited for prediction of short-term program engagement. The ROC curves are shown in Figure \ref{fig:short-term}. AUC scores of our models are much higher than that of random forest baseline.

\section{Long term drop-off prediction}

We next focus on the more ambitious goal of identifying women who are at a high risk of dropping out of the program based on their engagement behaviour in the first two months from registration. These predictions can help identify beneficiaries who require more involvement from health care workers, and plan for long-term interventions such as providing access to counselors. 

\paragraph{Problem Formulation}

Predicting program drop-off requires us to consider a longer period of time for generating labels, we generate labels that disentangle the effect of connection failure by considering the ratio of engagements to connections. Note that considering this ratio without setting constraints on the number of connections will introduce noise in the labels. After calculating the ratio, we define a risk threshold that helps us binarize the ratio into high risk and low risk. For example, if a beneficiary has engaged 5 out of the 15 times that the connection was successful, the engagements to connections ratio is 0.33 and on binarizing this with a risk threshold of 0.5, the beneficiary is labelled \texttt{high risk}.

We also consider the fraction of attempted calls that successfully connect in order to identify beneficiaries that are likely to disengage due to the small fraction calls they receive. Hence, we formulate an additional problem to identify beneficiaries that are less likely to connect regularly in the program. We generate labels by computing the connections to attempts ratio and using a risk threshold to binarize the data into low and high risk classes.

In the engagement risk classification, we consider beneficiaries that have atleast eight months of call logs available. We use the first two months for calculating input features and the rest of the call logs for generating the labels. We additionally require that a beneficiary has a minimum of 24 connections in the prediction period. Our final dataset contains data for 85,076 beneficiaries.

It is important to note that the value of risk threshold dictates the amount of imbalance in the classes. We use a risk threshold of 0.5 which results in 49,556 low risk beneficiaries and 35,520 high risk beneficiaries. We also learnt that a model with high recall for high risk beneficiaries would be more beneficial. Therefore, to account for the imbalance in the dataset and to prioritize high recall on high risk beneficiaries, we train models using different class weights for low risk and high risk classes. 

In the connection risk classification, we consider all the beneficiaries that have at least eight months of call logs available and at least 24 attempts were made. For this task, our dataset contains data for 162,301 women. We consider a risk threshold of 0.25 in this formulation. This means that only beneficiaries with a connections to attempts ratio of less than 0.25 will be considered \texttt{high risk}. This task, though significantly more challenging, can help identify beneficiaries that have very low likelihood of connection. The dataset for this task contains 83,455 low risk beneficiaries and 78,846 high risk beneficiaries.

\paragraph{Models}

We used the same models, training process and evaluation procedure as the previous formulation. In the CoNDiP architecture, we reduced the number of feed forward layers for encoding static features to 1 layer with 12 units. Each convolutional layer has 8 filters of size 3. There is 1 feed forward layer of 10 units after concatenating both the static and dynamic feature encodings. The same architecture is used for both engagement risk classification and connection risk classification. For the random forest, we used 100 trees with a maximum depth of 10. In the ReNDiP architecture, we encode the static features by using a feed foward layer of 12 units. We reduce the number of units to 8 in the LSTM layer. The concatenated static and dynamic encodings are fed to a feed forward layer of 10 units. We also limit the maximum number of calls considered in the input period to 18 (2 calls per week on average for 60 days). 

\begin{table}[h]
\centering
\begin{tabular}{|l|l|l|l|l|}
\hline
Model  & Accuracy & Precision & Recall & F1 score \\ \hline
RF     & 0.73     & 0.73      & 0.73   & 0.72     \\ \hline
CoNDiP & 0.75     & 0.75      & 0.75   & 0.75     \\ \hline
ReNDiP & 0.76     & 0.76      & 0.76   & 0.76     \\ \hline
\end{tabular}
\caption{Results for long-term engagement task, RF: Random Forest}
\label{tab:long-term-engagemnt}
\end{table}

\begin{table}[h]
\centering
\begin{tabular}{|l|l|l|l|l|}
\hline
Model  & Accuracy & Precision & Recall & F1 score \\ \hline
RF     & 0.69     & 0.69      & 0.70   & 0.69     \\ \hline
CoNDiP & 0.69     & 0.69      & 0.71   & 0.70     \\ \hline
ReNDiP & 0.70     & 0.69      & 0.71   & 0.70     \\ \hline
\end{tabular}
\caption{Results for long-term connection task, RF: Random Forest}
\label{tab:long-term-connection}
\end{table}

\begin{figure}[h]
    \centering
    \includegraphics[width=.8\linewidth]{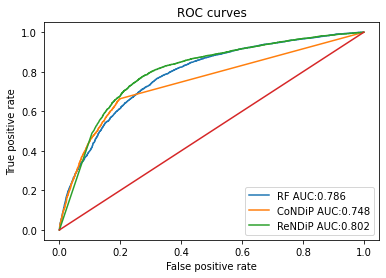}
    \caption{ROC curve and area under the curve (in legend) for the models for long-term engagement task}
    \label{fig:long-term-engagement}
\end{figure}

\begin{figure}[h]
    \centering
    \includegraphics[width=.8\linewidth]{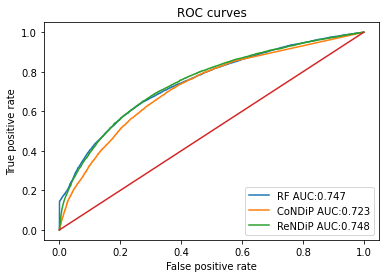}
    \caption{ROC curve and area under the curve (in legend) for the models for long-term connection task}
    \label{fig:long-term-connection}
\end{figure}

\paragraph{Results}

The results for engagement risk classification task and connection risk classification task are summarized in Table \ref{tab:long-term-engagemnt} and Table \ref{tab:long-term-connection} respectively. These models were trained with class weights of 1 and 1.5 for low risk and high risk classes respectively. For engagement risk classification, we see that ReNDiP and CoNDiP perform better than the baseline in terms of accuracy with ReNDiP performing the best. However, for the connection risk classification task, we see that all three models perform similarly. 

The ROC curves for both the tasks are shown in Figure \ref{fig:long-term-engagement} and Figure \ref{fig:long-term-connection}. ReNDiP's AUC is slighly better than the baseline whereas CoNDiP's is worse. It is clear that for long term prediction tasks, ReNDiP is the best suited.

\section{Deployment study}
We developed an easy-to-use dashboard for health workers to visualize predicted risk of drop-off. 
We piloted our long term drop-off prediction model and the dashboard app on beneficiaries registered in November 2019. This included 18,766 beneficiaries. However, only 17,288 beneficiaries had at least 8 attempts made to them in the first 60 days. 

Using ReNDiP to study the relevance of the predictions, we gathered the available call data for these beneficiaries, removed the call logs of the first 60 days (as they were used to build input features) and calculated the engagement to connection ratio for each beneficiary. To correctly analyse the performance of the model, we set a threshold for number of connections (to discount beneficiaries with fewer connection). The results are summarized in Table \ref{tab:pilot-overall}.

\begin{table}[h]
\centering
\begin{tabular}{|l|l|l|l|l|}
\hline
MC & Accuracy & Precision & Recall & F1 score \\ \hline
5  & 0.79     & 0.80      & 0.79   & 0.79     \\ \hline
10 & 0.81     & 0.82      & 0.81   & 0.81     \\ \hline
15 & 0.84     & 0.85      & 0.84   & 0.84     \\ \hline
\end{tabular}
\caption{Results for long-term engagement task in pilot study. MC stands for minimum connections.}
\label{tab:pilot-overall}
\end{table}

We consider thresholds of 5, 10 and 15 to understand how the model's predictions hold with different estimates of beneficiary behaviour. We observe that the model performs much better than it did during training phase, partly due to a change in the underlying data distribution. There was a greater proportion of low risk beneficiaries in the pilot study and as the model is better at predicting low risk beneficiaries, the model's performance becomes better. We also observe that the performance increases as we have better estimates of the beneficiaries behaviour. It is also important to note that low risk beneficiaries are more likely to have greater number of connections. This may explain the increasing performance with an increase in 'minimum connections' threshold. To ensure that the results are independent of any behavioural changes that may have been caused by SARS-CoV-2 pandemic, we separately analyse the call logs until 10 March 2020 as pre-pandemic period. The results for this analysis are summarized in Table \ref{tab:pilot-prepandemic}. 

\begin{table}[h]
\centering
\begin{tabular}{|l|l|l|l|l|}
\hline
MC & Accuracy & Precision & Recall & F1 score \\ \hline
5  & 0.81     & 0.81      & 0.81   & 0.81     \\ \hline
10 & 0.85     & 0.84      & 0.85   & 0.85     \\ \hline
\end{tabular}
\caption{Results for long-term engagement task in pre-pandemic period. MC stands for minimum connections.}
\label{tab:pilot-prepandemic}
\end{table}

\section{Discussion}

We have successfully built robust models to aid health care workers in NGOs such as ARMMAN to identify women at risk of dropping out of call-based health awareness service. Leveraging call history data collected from the past we have built deep-learning models that provide 85\%  and 76\% recall for short term and long term prediction tasks respectively.

As part of our future work, we aim to find the important set of features that lead to the predictions that our model outputs using interpretation techniques like \cite{selvaraju2017grad, merrick2019explanation}. We would utilize data from effects of ad-hoc interventions based on our model to help predict the right intervention or sequence of interventions  using reinforcement learning approaches.




\section{Acknowledgements}

We thank Jahnvi Patel and Yogesh Tripathi for their assistance with analysing the data. This research was supported by Google AI for Social Good faculty award to Balaraman Ravindran, (Ref: RB1920CP928GOOGRBCHOC).

\printbibliography 

@article{HUANG20121414,
title = "Customer churn prediction in telecommunications",
journal = "Expert Systems with Applications",
volume = "39",
number = "1",
pages = "1414 - 1425",
year = "2012",
author = "Bingquan Huang and Mohand Tahar Kechadi and Brian Buckley"
}

@article{HUNG2006515,
title = "Applying data mining to telecom churn management",
journal = "Expert Systems with Applications",
volume = "31",
number = "3",
pages = "515 - 524",
year = "2006",
author = "Shin-Yuan Hung and David C. Yen and Hsiu-Yu Wang"
}

@article{XIE20095445,
title = "Customer churn prediction using improved balanced random forests",
journal = "Expert Systems with Applications",
volume = "36",
number = "3, Part 1",
pages = "5445 - 5449",
year = "2009",
author = "Yaya Xie and Xiu Li and E.W.T. Ngai and Weiyun Ying"
}

@INPROCEEDINGS{5284154,  author={J. {Kawale} and A. {Pal} and J. {Srivastava}},  booktitle={2009 International Conference on Computational Science and Engineering},   title={Churn Prediction in MMORPGs: A Social Influence Based Approach},   year={2009},  volume={4},  number={},  pages={423-428},}

@inproceedings{runge2014churn,
  title={Churn prediction for high-value players in casual social games},
  author={Runge, Julian and Gao, Peng and Garcin, Florent and Faltings, Boi},
  booktitle={2014 IEEE conference on Computational Intelligence and Games},
  pages={1--8},
  year={2014},
  organization={IEEE}
}

@ARTICLE{10.3389/fphar.2013.00091,
AUTHOR={Kardas, Przemyslaw and Lewek, Pawel and Matyjaszczyk, Michal},
TITLE={Determinants of patient adherence: a review of systematic reviews},
JOURNAL={Frontiers in Pharmacology},
VOLUME={4},
PAGES={91},
YEAR={2013}
}

@article{son2010application,
  title={Application of support vector machine for prediction of medication adherence in heart failure patients},
  author={Son, Youn-Jung and Kim, Hong-Gee and Kim, Eung-Hee and Choi, Sangsup and Lee, Soo-Kyoung},
  journal={Healthcare informatics research},
  volume={16},
  number={4},
  pages={253--259},
  year={2010}
}

@inproceedings{howes2012predicting,
  title={Predicting adherence to treatment for schizophrenia from dialogue transcripts},
  author={Howes, Christine and Purver, Matthew and McCabe, Rose and Healey, Patrick GT and Lavelle, Mary},
  booktitle={Proceedings of the 13th Annual Meeting of the Special Interest Group on Discourse and Dialogue},
  pages={79--83},
  year={2012},
  organization={Association for Computational Linguistics}
}

@article{HIV,
    author = {Tuldrà, Albert and Ferrer, Ma José and Fumaz, Carmina R. and Bayés, Ramon and Paredes, Roger and Burger, David M. and Clotet, Bonaventura},
    title = "{Monitoring Adherence to HIV Therapy}",
    journal = {Archives of Internal Medicine},
    volume = {159},
    number = {12},
    pages = {1376-1377},
    year = {1999},
    month = {06},
}

@article{10.1001/archinte.1996.00440020063008,
    author = {Pilote, Louise and Tulsky, Jacqueline P. and Zolopa, Andrew R. and Hahn, Judith A. and Schecter, Gisela F. and Moss, Andrew R.},
    title = "{Tuberculosis Prophylaxis in the Homeless: A Trial to Improve Adherence to Referral}",
    journal = {Archives of Internal Medicine},
    volume = {156},
    number = {2},
    pages = {161-165},
    year = {1996},
    month = {01}
}

@article{Killian_2019,
   title={Learning to Prescribe Interventions for Tuberculosis Patients Using Digital Adherence Data},
   journal={Proceedings of the 25th ACM SIGKDD International Conference on Knowledge Discovery \& Data Mining},
   publisher={ACM},
   author={Killian, Jackson A. and Wilder, Bryan and Sharma, Amit and Choudhary, Vinod and Dilkina, Bistra and Tambe, Milind},
   year={2019},
   month={Jul}
}

@TECHREPORT{10665-327596,
	author = {World Health Organization},
	title = {Trends in maternal mortality 2000 to 2017: estimates by WHO, UNICEF, UNFPA, World Bank Group and the United Nations Population Division: executive summary},
	year = {2019},
	pages = {12 p.},
	publisher = {World Health Organization},
	type = {Technical documents}
}

@article{thaddeus1994too,
  title={Too far to walk: maternal mortality in context},
  author={Thaddeus, Sereen and Maine, Deborah},
  journal={Social science \& medicine},
  volume={38},
  number={8},
  pages={1091--1110},
  year={1994},
  publisher={Pergamon}
}

@article{kiranyaz20191d,
  title={1D convolutional neural networks and applications: A survey},
  author={Kiranyaz, Serkan and Avci, Onur and Abdeljaber, Osama and Ince, Turker and Gabbouj, Moncef and Inman, Daniel J},
  journal={arXiv preprint arXiv:1905.03554},
  year={2019}
}

@article{hochreiter1997long,
  title={Long short-term memory},
  author={Hochreiter, Sepp and Schmidhuber, J{\"u}rgen},
  journal={Neural computation},
  volume={9},
  number={8},
  pages={1735--1780},
  year={1997},
  publisher={MIT Press}
}

@inproceedings{selvaraju2017grad,
  title={Grad-cam: Visual explanations from deep networks via gradient-based localization},
  author={Selvaraju, Ramprasaath R and Cogswell, Michael and Das, Abhishek and Vedantam, Ramakrishna and Parikh, Devi and Batra, Dhruv},
  booktitle={Proceedings of the IEEE international conference on computer vision},
  pages={618--626},
  year={2017}
}

@article{merrick2019explanation,
  title={The Explanation Game: Explaining Machine Learning Models with Cooperative Game Theory},
  author={Merrick, Luke and Taly, Ankur},
  journal={arXiv preprint arXiv:1909.08128},
  year={2019}
}

\end{document}